\documentclass[aps,prb,twocolumn,groupedaddress]{revtex4}

\usepackage{graphics} 
\usepackage{graphicx}
\usepackage{dcolumn}
\usepackage{bm}
\everymath{\displaystyle}
\usepackage{amsmath}

\setcitestyle{super}

\begin{document}

\title{Magneto-intersubband resistance oscillations  in GaAs quantum wells\\ placed in a tilted magnetic field.} 

\author{William Mayer}
\author{Jesse Kanter}
\author{Javad Shabani}
\author{Sergey Vitkalov}
\email[Corresponding author: ]{vitkalov@sci.ccny.cuny.edu}
\affiliation{Physics Department, City College of the City University of New York, New York 10031, USA}
\author{A. K. Bakarov} 
\affiliation{A.V.Rzhanov Institute of Semiconductor Physics, Novosibirsk 630090, Russia}
\author{A. A. Bykov}
\affiliation{A.V.Rzhanov Institute of Semiconductor Physics, Novosibirsk 630090, Russia}
\affiliation{Novosibirsk State University, Novosibirsk 630090, Russia}

\date{\today}

\begin{abstract} 

The magnetotransport of highly mobile 2D electrons in wide GaAs single quantum wells with three populated subbands placed in  titled magnetic fields is studied. The bottoms of the lower two subbands have nearly the same energy while the bottom of the third subband has a much higher energy ($E_1\approx E_2<<E_3$). At zero in-plane magnetic fields magneto-intersubband oscillations (MISO) between the $i^{th}$ and $j^{th}$ subbands are observed and obey the relation $\Delta_{ij}=E_j-E_i=k\cdot\hbar\omega_c$, where $\omega_c$ is the cyclotron frequency and $k$ is an integer. An application of  in-plane magnetic  field produces dramatic  changes in MISO and the corresponding electron spectrum. Three regimes are identified. At $\hbar\omega_c \ll \Delta_{12}$ the in-plane magnetic field increases considerably the gap $\Delta_{12}$, which is consistent with the semi-classical regime of   electron propagation. In contrast at strong magnetic fields  $\hbar\omega_c \gg \Delta_{12}$   relatively weak oscillating variations of the electron spectrum with the in-plane magnetic field are observed. At $\hbar\omega_c \approx \Delta_{12}$ the electron spectrum undergoes a transition between  these two  regimes through  magnetic breakdown. In this transition regime MISO with odd quantum number $k$ terminate, while MISO corresponding to  even $k$ evolve $continuously$ into the high field regime corresponding to  $\hbar\omega_c \gg \Delta_{12}$.        

\end{abstract}
 

\maketitle

\section{Introduction}

The quantization of electron motion in magnetic fields generates a great variety of  fascinating phenomena observed in condensed materials. A well-known example is the Shubnikov-de Haas (SdH) resistance oscillations\cite{shoenberg1984}. The passage of strongly degenerate Landau levels through the Fermi surface at a low temperature $T$ produces resistance oscillations due to a modulation of the net number of electron states in the energy interval $kT<\hbar \omega_c$ near the Fermi energy $E_F$ that provide the dominant contribution to electron transport \cite{ando1982,ziman}. In two dimensional electron systems, SdH oscillations can be very pronounced \cite{ando1982}, leading to the Quantum Hall Effect (QHE) at low temperatures $kT \ll \hbar \omega_c$ \cite{qhe}. 

Landau quantization  produces a remarkable  effect on  Joule heating  of two dimensional (2D) electrons \cite{dmitriev2005,vitkalov2007,zhang2009,mamani2009}.  The heating forces 2D electrons into exotic electronic states in which voltage (current) does not depend on current \cite{bykov2007zdr,zudov2008zdr,gusev2011zdr} (voltage\cite{bykov2013zdc}). In contrast to the linear response at low temperatures $kT < \hbar \omega_c$ (SdH, QHE),  the quantization affects Joule heating  in a significantly broader temperature range. At  $kT \gg \hbar \omega_c$ the $dc$ heating produces a multi-tiered electron distribution containing as many tiers as the number of Landau levels inside the energy interval $kT$: $N \approx kT/\hbar \omega_c$. This quantal heating preserves the overall broadening ($\sim kT$) of the electron distribution \cite{zhang2009,romero2008}.  Surprisingly the electron distribution resulting from quantal heating is, in some respect, similar to the one created by  quantum microwave pumping  between Landau levels\cite{zudov2001,ye2001}. Indicated phenomena produce a broad variety of  nonlinear effects   in quantizing magnetic fields and present an exciting area of  contemporary research\cite{dmitriev2012}. 
 
Two-dimensional electron systems with multiple populated subbands exhibit additional  quantum magnetoresistance oscillations\cite{coleridge1990,leadley1992,bykov2008a,bykov2008b,bykov2008c,mamani2008,bykov2009,goran2009}.  These magneto-inter-subband oscillations (MISO) of the resistance  are due to an alignment between Landau levels from different subbands $i$ and $j$ with corresponding energies $E_i$ and $E_j$. Resistance maxima occur at magnetic fields at which the gap between the bottoms of subbands, $\Delta_{ij}=E_i-E_j$, equals a multiple of the Landau level spacing, $\hbar \omega_c$: $\Delta_{ij}=k\cdot\hbar\omega_c$, where $k$ is an integer \cite{magarill1971,polyanovskii1988,raikh1994,raichev2008}.  At this condition  electron  scattering on rigid impurities is enhanced due to  the possibility of electron transitions between  $i^{th}$ and $j^{th}$ subbands. At magnetic fields corresponding to the condition $\Delta_{ij}=(k+1/2)\cdot\hbar\omega_c$ the intersubband electron transitions are suppressed. As a result, the resistance oscillations are periodic in inverse magnetic field due to this modulation of electron scattering. In contrast to SdH oscillations MISO are less sensitive to  temperature. MISO are observed at  temperatures, $kT \gg \hbar \omega_c$ at which SdH oscillations (and QHE) are absent. 

This paper presents investigations of MISO in  wide quantum wells with three populated subbands placed in a tilted magnetic field. Studied systems contain conducting electrons  localized near  the edges of the quantum wells. The electrons,  thus, form two parallel 2D  systems separated by a distance $d$.   A weak electron  tunneling between these  two systems occurs through a relatively wide but shallow potential. In zero magnetic field the lateral (along 2D systems) and vertical (between 2D systems) motions of an electron are completely disentangled.  The vertical tunneling  uniformly splits the  electron spectrum originally degenerate in the lateral directions. The resulting eigenvalues correspond to symmetric ($E_1$) and antisymmetric ($E_2$) configurations of electron wave functions in the vertical direction with the energy gap $\Delta_{12}=E_2-E_1$ between two subbands independent of the lateral wave vector $\vec k$. 

The bottom of the third subband  has a much higher electron energy $E_3 \gg E_{1,2}$. Application of  perpendicular magnetic field quantizes the lateral motion in all subbands  inducing MISOs. The MISOs corresponding to electron scattering between the third  and the two lower subbands oscillate at  high frequencies and demonstrate a distinct beating pattern. This useful property provides a very accurate measurement of the evolution of  the electron spectrum in response to  in-plane magnetic field.

Application of  in-plane magnetic field couples the symmetric and antisymmetric states\cite{bobinger1991,hu1992} leading to  a significant modification of the electron spectrum. Theoretical investigations of the electronic structure of two parallel  2D dimensonal electron systems in tilted magnetic fields have revealed three regimes occurring at   small  ($\hbar\omega_c \ll \Delta_{12}$, semi-classical (SC) regime), strong  ($\hbar\omega_c \gg \Delta_{12}$, high field (HF) regime) and intermediate  (magnetic breakdown (MB) regime) magnetic fields\cite{hu1992}. The intermediate magnetic fields   correspond to  magnetic breakdown\cite{priestley1963,cohen1961,blount1962,slutskin1968} of the semi-classical electron spectrum leading generally to complex combinations of semi-classical orbits \cite{pippard1962,pippard1964,slutskin1968}. These regimes  have been investigated  to a different extent mainly in double quantum wells using SdH oscillations and  QHE.\cite{bobinger1991,harff1997,mueed2015,gusev2007,gusev2008} However the results have not been  compared coherently with the theory accross  all three regimes and  important properties of the quantum oscillations have not been revealed.  We note also that in the regime of QHE the  energy spectrum  is often sensitive to effects of the electrostatic redistribution of 2D carriers between different subbands and quantum levels, which makes a quantitative comparison between different regimes  challenging \cite{kuk2009,smet2012}. 
  
This paper presents an attempt to study the evolution of the electron spectrum in wide quantum wells with  in-plane magnetic field using MISO. The  experiments are performed at a high temperature $kT \gg \hbar\omega_c$ at which effects of the  electrostatic electron redistribution between Landau levels are, most likely, not relevant. The paper shows both MISO and SdH oscillations obtained at different angles $\alpha$ between the magnetic field and the normal to the 2D systems and yields  a detailed evolution of electron spectrum in multi-subband 2D systems in a broad range of magnetic fields.  Presented results  show a termination of the MISO corresponding to $\Delta_{12} =k\cdot \hbar \omega_c$ at odd $k$ in the magnetic breakdown regime. The termination is accompanied by a collapse of the nodes in the beating between MISOs corresponding to the third subband.   The obtained data demonstrate  a good agreement with numerical simulations  based on the existing theory\cite{hu1992} in a broad range of magnetic fields including all regimes indicated above. 

Our experiments have revealed  an outstanding sensitivity of the electron spectrum to the angle  $\alpha$, especially at  $\hbar\omega_c \approx \Delta_{12}$. The sensitivity to  in-plane magnetic field  is due to both  a strong Lorent's force, which occurs in  studied samples with high electron density, and a weak tunneling  between 2D parallel systems. The presented results indicate  that the recently observed ambiguity in the MISO amplitude at $k$=1  is, most likely, related to a small misalignment between the direction of the magnetic field and the normal to  2D sample in different measurements \cite{wiedmann2010,dietrich2015}.

\section{Experimental Setup}

Studied GaAs quantum wells were grown by molecular beam epitaxy on a semi-insulating (001) GaAs substrate. The material was fabricated from a selectively doped GaAs single quantum well of width $d=$56 nm sandwiched between AlAs/GaAs superlattice barriers. The heterostructure has three populated subbands with energies $E_1\approx E_2<<E_3$ at the bottoms of the subbands. The subband energies are  schematically shown in the insert to Figure \ref{data}.

The studied samples were etched in the shape of a Hall bar. The width and the length of the measured part of the samples are $W=50\mu$m and $L=250\mu$m. AuGe eutectic was used to provide electric contacts to the 2D electron gas. Two samples were studied at temperature 4.2 Kelvin in magnetic fields up to 4 Tesla applied $in$-$situ$ at different angle $\alpha$  relative  to the normal to 2D layers and  perpendicular to the applied current.  The angle $\alpha$ has been evaluated using Hall voltage $V_H = B_\perp/(en_T)$, which is proportional to the perpendicular component, $B_\perp=B\cdot cos(\alpha)$, of the total magnetic field $B$.     The total electron density of samples, $n_T\approx 8.6\times 10^{11} cm^{-2}$, was evaluated from the Hall measurements taken at $\alpha$=0$^0$ in classically strong magnetic fields \cite{ziman}. An average electron mobility $\mu \approx 1.6 \times 10^6 cm^2/Vs$  was obtained from $n_T$ and the zero-field resistivity.  Sample resistance was measured using the four-point probe method. We applied a 133 Hz $ac$ excitation $I_{ac}$=1$\mu$A  through the current contacts and measured the longitudinal and Hall $ac$ voltages ($V^{ac}_{xx}$ and $V^{ac}_H$) using two lockin amplifiers with 10M$\Omega$ input impedances. The potential contacts provided insignificant contribution to the overall response due to small values of the contact resistance (about 1k$\Omega$) and negligibly small electric current flowing through the contacts. The measurements were done in the linear regime in which the voltages are proportional to the applied current.

\section{Results and Discussion}

A theoretical analysis  yields the following expression for the amplitude of  MISO due to the scattering between the $i^{th}$ and $j^{th}$ subbands in  weak ($\omega_c \tau_q^{(i)}<1$) perpendicular magnetic fields \cite{raikh1994,raichev2008}:

\begin{align}
\Delta\rho_{MISO}^{(i,j)}=&\frac{2m\cdot\nu_{ij}}{e^2(n_i+n_j)}\cdot \cos\left(\frac{2\pi\Delta_{ij}}{\hbar\omega_c}\right) \nonumber\\ 
&\times \exp\left[\frac{-\pi}{\omega_c}\left(1/\tau_q^{(i)}+1/\tau_q^{(j)}\right)\right],
\label{miso}
\end{align}
where  $n_i$ and $\tau_q^{(i)}$ are  the electron density and quantum scattering time\cite{vavilov2004}  in the $i^{th}$ subband, $\nu_{ij}$ is an effective intersubband transport scattering rate,  $m$ is the effective electron mass and $\omega_c=eB_\perp/m$ is the cyclotron frequency \cite{raichev2008}. This expression has recently been used   in systems with two and three populated subbands to extract the quantum  scattering rate $1/\tau_q^{i}$  \cite{bykov2008a,bykov2008b,bykov2008c,mamani2008,bykov2009,goran2009,wiedmann2010,dietrich2015}. The expression indicates that  MISO between $i^{th}$ and $j^{th}$ subbands are periodic in  inverse magnetic field $1/B_\perp$.

\begin{figure}[t]
\vskip -0.4cm
\includegraphics[width=\columnwidth]{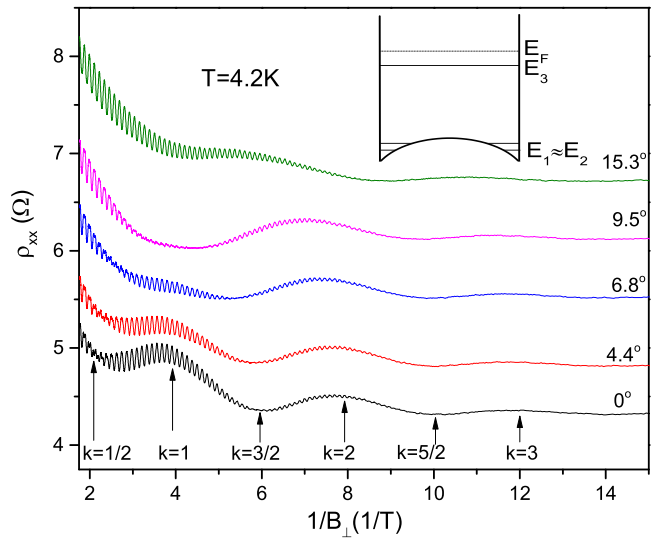}
\caption{(Color online) Dependencies of the longitudinal resistance $\rho_{xx}$ on the inversed component of the magnetic field, which is perpendicular to the 2D sample, $1/B_\perp$,  obtained at different angles $\alpha$ between the total magnetic field $\vec B$ and the normal to the samples as labeled. Integer values of index $k$ corresponds to the  maximums of LF-MISO at $\Delta_{12}=k \cdot \hbar \omega_c$ (see Eq.(\ref{miso})) and to anti-nodes of the  beat pattern of HF-MISO.  Half-integer value of $k$  corresponds to the minimums of LF-MISO and the nodes of the HF-MISO beat pattern at angle $\alpha=0^0$. Sample A. The insert presents the  energy diagram of studied samples.  }
\label{data}
\end{figure}

Figure \ref{data} presents the longitudinal  resistivity, $\rho_{xx}(1/B_\perp)$, of sample A  at different angles $\alpha$ between the magnetic field and the normal to the sample as labeled. At $\alpha=0^0$ in accordance with Eq.(\ref{miso}) the frequency of MISO  in inverse magnetic field  is proportional to the intersubband energy gap ($f_{ij}\propto\Delta_{ij}=E_i-E_j$). This three subband system should therefore have MISOs at three different frequencies, corresponding to resonant scattering between the  subbands. MISOs associated with scattering between the two lowest  subbands  have a low frequency (LF-MISO), $f_{21} \propto E_2-E_1$, since the energy gap $\Delta_{21}$ is very small ($E_1 \approx E_2$). The two sets of MISOs associated with scattering between the upper band and each of the lower bands  have much higher frequencies, $f_{31} \approx f_{32} \gg f_{12}$ (HF-MISO), that are approximately equal since $\Delta_{31} \approx \Delta_{32} \gg \Delta_{21}$. Due to the small difference between energy $E_1$ and $E_2$ the interference between  MISOs with frequencies $f_{31}$ and  $f_{32}$  produces a beating pattern with a small beating frequency $f_{beat} \propto (E_2-E_1)/2 \ll f_{3i}$ and a high inner frequency $f_+\propto (2E_3-E_2-E_1)/2$. The  resistance oscillations with  both the low ($f_{21}$) and high ($f_{31},f_{32}$) frequencies   are shown in Figure \ref{data}. 

\begin{figure}[t]
\includegraphics[width=3.4in]{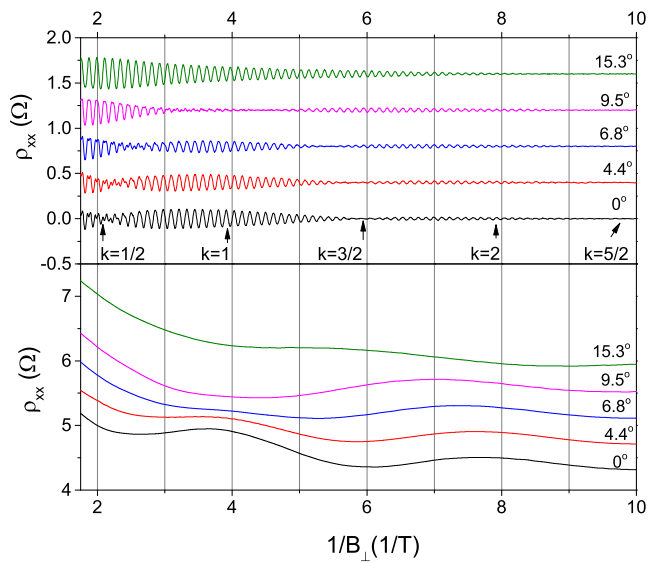}
\caption{(Color online) Lower (upper) panel presents LF-MISO (HF-MISO) obtained by a low (high) frequency  FFT filtering of  the magnetoresitance oscillations presented in Fig.\ref{data}.  Sample A.}
\label{smooth}
\end{figure}

The significant frequency difference between the low and high frequency contents of oscillations  facilitates the separation of HF and LF-MISOs by  fast Fourier transform (FFT) filtering. In Figure \ref{smooth} the lower panel presents the low frequency content  while the upper panel presents the high frequency oscillations, which have been filtered from the curves presented in Fig.\ref{data}\cite{dietrich2015}.  

\begin{figure}[t]
\includegraphics[width=0.94\columnwidth]{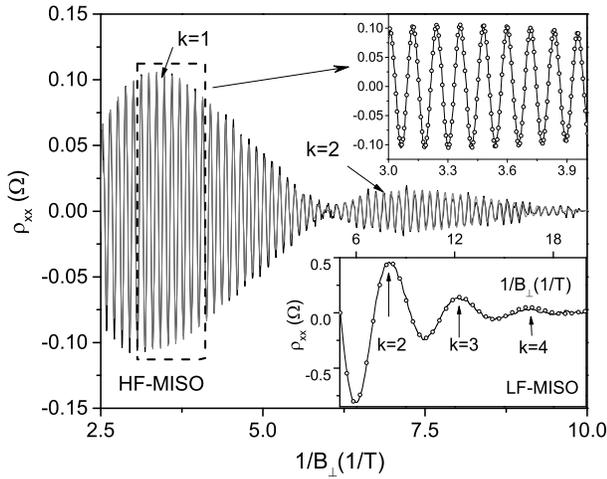}
\caption{Comparison of HF-MISO shown by the black solid line with the theoretical dependence  based on  Eq.(\ref{miso}) and  shown by the gray line. Upper insert demonstrates  a more detailed view of the comparison.  Lower insert show  a comparison of  LF-MISO with the theory. In both inserts  open circles present  theoretical dependencies. Sample A.}
\label{mis_teo}
\end{figure} 

Due to the precise relation between different frequencies the beating frequency is half of  the frequency of MISO corresponding to the two lower subbands: $f_{beat}=f_{21}/2$ at $\alpha$=0$^0$. This is indeed seen in Figure \ref{smooth}.  Figures \ref{data} and \ref{smooth} show that at $\alpha$=0$^0$ the nodes of HF-MISO correspond to the minimums of the LF-MISO. Furthermore an analysis of the HF-MISO phase indicates that   in the  $k$=2 region the phase of HF-MISO is shifted by $\pi$ with respect to the HF-MISO phase in $k$=1 region at $\alpha$=0$^0$. To verify this $\pi$-phase shift, we compare HF-MISO at $\alpha$=0$^0$ with the one at  $\alpha$=15.3$^0$, which demonstrates  no nodes and is perfectly periodic with respect to $1/B_\perp$.  The comparison shows that in the  $k$=1 region the maximums of HF-MISO at $\alpha$=0$^0$ corresponds to the minimums of HF-MISO at $\alpha$=15.3$^0$, while in $k$=2 region  the maximums of HF-MISO at $\alpha$=0$^0$ corresponds to the maximums of HF-MISO at $\alpha$=15.3$^0$.   Thus  the observed interference of  HF-MISOs at $\alpha$=0$^0$ corresponds to the beating between two frequencies at $f_{beat}=f_{21}/2$.

 Figure \ref{mis_teo} demonstrates the direct comparison of HF-MISO  at $\alpha$=0$^0$ with Eq.(\ref{miso}). The experiment  agrees well with the theory in the whole range of magnetic fields corresponding to $\Delta_{12}>\hbar \omega_c$. At higher magnetic fields a quantitative comparison has not been accomplished  due to the presence of SdH oscillations and higher harmonics of MISO, which are not captured in Eq.(\ref{miso}).    Figure \ref{mis_teo}  presents  also a comparison of LF-MISO with the theory. Shown in Fig.\ref{data} the monotonic background  corresponding to the positive quantum magnetoresistance\cite{vavilov2004,dietrich2012} has been removed by a procedure reported earlier \cite{dietrich2015}. However  the relatively strong increase of the resistance observed at  $\hbar \omega_c>\Delta_{12}$ interferes with the low frequency oscillating content making the applied procedure to be quite uncertain  there.  This comparison is limited to  the magnetic fields corresponding to  condition $k>$3/2. In this range of magnetic fields a very good agreement between LF-MISO and the theory is found. A joint  analysis of LF-MISO and HF-MISO yields the quantum scattering time in each subband \cite{dietrich2015}.  In lower subbands  the time is found to be $\tau_q^{(1,2)}$=8.2 $\pm 0.3$ ps   while in the third subband the time is $\tau_q^{(3)}$=3.5 $\pm 0.3$ ps. These  values agree with those ones obtained  in similar systems with three populated subbands \cite{dietrich2015}.

\begin{figure}[t]
\includegraphics[width=0.80\columnwidth]{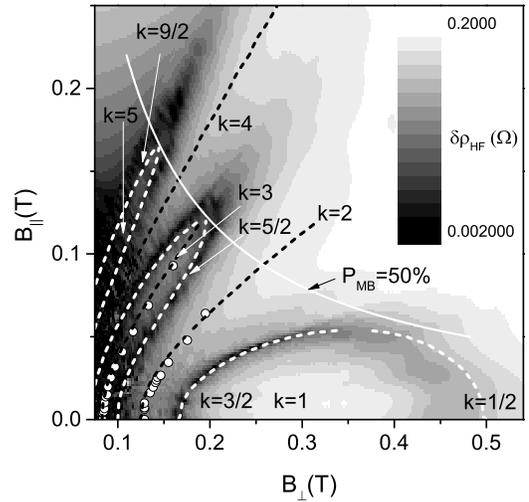}
\caption{Dependence of HF-MISO magnitude on $B_\perp$ and $B_\parallel$. Black color presents locations of HF-MISO nodes.  Open circles present  experimental positions of LF-MISO maximums.  White (black) dash lines present position of HF-nodes (LF-maximums) obtained using numerical calculations of electron spectrum. White solid line corresponds to  50\% probability of  magnetic breakdown of semi-classical trajectories\cite{hu1992}. All spectra are obtained at $t_0$=0.215 meV and $d$=36 nm. Size of the circles corresponds to experimental uncertainty of the position. Sample A.}
\label{2D}
\end{figure} 

An introduction of a parallel magnetic field, $B_\parallel$,  produces  significant changes in MISO. The most notable is the disappearance of the $k$=1 maximum, which occurs near angle $\alpha$=9.5$^0$ in Fig.\ref{smooth}. This disappearance is accompanied by a spectacular collapse of two nodes of HF-MISO corresponding to $k$=1/2 and $k$=3/2 at $\alpha$=0$^0$. These nodes collapse in the vicinity of the main LF-MISO maximum $k$=1.   

Figure \ref{2D} presents the evolution of the magnitude of  HF-MISO in the ($B_\perp$-$B_\parallel$) plane. The HF-MISO magnitude (the envelop of HF-MISO) 
was obtained by a low frequency  filtering of the square of  HF-MISO: 
$\delta \rho_{HF}= (2\langle A^2 cos^2(2\pi f_{+}/B_\perp)\rangle)^{1/2}=A$, 
where $A(B_\perp,B_\parallel)$ is the slowly varying magnitude of HF-MISO and angle brackets stand for the low frequency filtering. The  low pass filter rejects the high frequency content  of the squared HF-MISO but passes slow oscillations at the beating frequency. The applied procedure yields the envelop of the HF-MISO with  a standard deviation within  $\pm$0.004$\Omega$\cite{deviation}.  

Figure \ref{2D} shows very different behavior of the odd and even MISO maximums in  response to $B_\parallel$. The even  ($k$=2,4...) maximums of the MISO magnitude  evolve continuously  into the high magnetic field region, whereas the odd  ($k$=1,3...) maximums terminate within the  regions bounded  by  HF-MISO nodes as shown  in Figure \ref{2D}.  A transition from an odd region  to an even  region changes the phase of HF-MISO by $\pi$. 

The figure demonstrates an additional interesting MISO property in the $B_\perp$-$B_\parallel$ plane:  a possibility of  the continuous (without intersection with a node line)  transition between even maximums. Indeed  by an appropriate choice of the $B_\perp$ and $B_\parallel$ the MISO maximum at $k$=2 can be transfered into $k$=4 MISO maximum at $\alpha$=0$^0$ without intersecting the nodal lines. In this sense all even maximums are topologically equivalent. This set also includes  the $k$=0 maximum  corresponding to the limit of strong magnetic fields. In contrast an odd MISO maximum presents an energy spectrum, which is  topologically different from the spectrum corresponding to  strong magnetic fields. The latter  is the spectrum of  uncoupled 2D systems\cite{shoenberg1984,hu1992}.

\subsection{Numerical analysis of electron spectrum}
The evolution of  MISO with both in-plane and perpendicular magnetic fields  is found to be in   good agreement with numerical evaluations of the electron spectrum in those field\cite{axes}.  In this section we present a theory describing the effect of  in-plane magnetic field  on the electron spectrum of two 2D parallel electron systems\cite{hu1992,kumada2008}.  The theory treats the interlayer hopping in a tight binding approximation so that the single-particle problem is characterized by the interlayer distance  $d$ and hopping integral $t_0$\cite{hu1992}. In the titled magnetic field $\vec B=(-B_\parallel, 0, B_\perp)$ electrons are described by the Hamiltonian:
\begin{equation}
H=\frac{\hbar^2 k_x^2}{2m^*}+\frac{e^2B_\perp^2}{2m^*}x^2+\frac{\hbar^2 k_z^2}{2m^*}+V(z) 
+\frac{e^2B_\parallel^2}{2m^*}z^2+\frac{e^2B_\perp B_\parallel}{m^*}xz,\\
\label{ham}
\end{equation}
where $m^* $ is effective mass and $V(z)$ is the electrostatic potential between two 2D systems.  
To obtain Eq.(\ref{ham}) we have used the  gauge (0,$B_\perp x+  B_\parallel z$,0) of the vector potential  and applied the transformation $x \rightarrow x-\hbar k_y/eB_\perp$.  

The first four terms describe the coupled 2D electron systems  in a perpendicular magnetic field.  The corresponding eigenfunctions of the system are $\vert N,\xi\rangle$, where $N$=0,1,2.. presents  $N$-$th$ Landau level (the lateral quantization) and $\xi=S, AS$ describes the symmetric (S) and antisymmetric (AS) configurations of the wave function in the $z$-direction (vertical quantization).   Using  functions $\vert N,\xi\rangle$ as the basis set , one can present the Hamiltonian in  matrix form.  The matrix contains four  matrix blocks: $\hat H=(\hat E^S,\hat T; \hat T, \hat E^{AS})$, where the semicolon separates rows.The diagonal matrices, $\hat E^S$ and $\hat E^{AS}$, represent  energy of the symmetric and antisymmetric wave functions  in different orbital states $N$:    
\begin{equation}
E^{S,AS}_{mn}=\delta_{mn}[\hbar \omega_c((n-1)+\frac{1}{2}) \pm t_0+\frac{e^2B_\parallel^2d^2}{8m^*}],
\label{diag}
\end{equation}
where sign $-$($+$) corresponds to symmetric (antisymmetric) states and  indexes $m$=1,2...$N_{max}$ and $n$=1,2...$N_{max}$ numerate rows and columns of the matrix correspondingly. These indexes are related to the orbital number $N$: $n,m=N+1$, since the orbital number $N=0,1,2..$. In  numerical computations the maximum number $N_{max}$ is chosen  to be about twice larger than the orbital number $N_F$ corresponding to Fermi energy $E_F$. Further increase of $N_{max}$ show a very small (within 1\%) deviation from the dependencies obtained at $N_{max}\approx 2N_F$. 

The first term in Eq.(\ref{diag}) describes the orbital quantization of electron motion  while
the second term relates to  the electron tunneling between 2D layers. The shape of the wave function in the $z$-direction ($\xi(z)$) is determined by the third and fourth terms in Eq.(\ref{ham}). Due to the complete disentanglement between the vertical ($z$) and lateral motions at $B_\parallel$=0T the second term does not depend on $N$.  The tunneling  term reads: $\langle\xi\vert V(z)\vert \xi\rangle=\pm t_0$.  In the tight binding approximation $t_0$ is considered to be independent of $B_\parallel$\cite{hu1992}. As shown below this approximation  provides very good agreement with experiment.   The last term in Eq.(\ref{diag}) describes diamagnetic shift of the quantum levels and is related to the fifth term in Eq.(\ref{ham}). In the basis set $\vert N,\xi\rangle$ the diamagnetic  term is proportional to $\langle\xi\vert z^2\vert \xi\rangle=(d/2)^2$, since in the tight binding approximation the thin 2D layers are located at distance $z=\pm d/2$ from the origin of $z$ axes. The diamagnetic term does not depend on $N$.    

The off-diagonal matrix $\hat T$ is related to the last term in Eq.(\ref{ham}), which mixes symmetric and antisymmetric states. Since $x=l_{B\perp}(a^*+a )/\sqrt2$ works as the raising $a^*$ and lowering $a$ operators of the Landau orbits, the last term in Eq.(\ref{ham}) couples Landau levels  with   orbital numbers different by one. Here  $l_{B\perp}=(\hbar/eB_\perp)^{1/2}$ is the  magnetic length in $B_\perp$. As a result, for $n>m$ the matrix element $T_{mn}$   between states $\vert N, S\rangle$ and $\vert N+1, AS\rangle$ is   
\begin{equation}
\begin{split}
T_{mn}&=\delta_{m+1,n}\frac{e^2B_\parallel B_\perp l_{B\perp}}{m^*}\langle N\vert \frac{a^*+a}{\sqrt2}\vert N+1\rangle \langle S\vert z\vert AS\rangle\\
&=\delta_{m+1,n}\hbar \omega_c\Big[ \frac{B_\parallel d}{2B_\perp l_{B\perp}}\Big](n/2)^{1/2}
\end{split}
\label{offdiag}
\end{equation}
The matrix $\hat T$ is  a symmetric matrix: $T_{mn}=T_{nm}$. The Hamiltonian $\hat H$ is diagonalized numerically at different magnetic fields $B_\perp$ and $B_\parallel$. To analyze the spectrum the obtained eigenvalues of the Hamiltonian are numerated in ascending order using positive integer index $l$=1,2.... The electron transport depends on the distribution of the quantum levels in the interval $kT$ near the Fermi energy $E_F$\cite{ziman}.   Below we focus on this part of the spectrum.

In accordance with Eq.(\ref{miso}) a HF-node  corresponds to an equal separation ($\hbar \omega_c/2$) between nearest quantum levels  in the vicinity of Fermi energy, whereas a HF-anti-node occurs when the two nearest  levels coincide with each other, thus,  the energy separation between pairs of coinciding levels is $\hbar \omega_c$. We note, that,  in contrast to the nodes of HF-MISO, the positions of the maximums of the magnitude of HF-MISO  and maximums of LF-MISO, shown in Fig.\ref{data} and  Fig.\ref{mis_teo},  are affected by the Dingle factor  and, therefore, do not exactly correspond to the magnetic fields at which two nearest Landau levels coincides.   In accordance with Eq.(\ref{miso})  the exponential decrease of the Dingle factor reduces significantly the beating magnitude at small magnetic fields and, thus,  shifts  the maximums of the beating pattern  to   higher magnetic fields.  Figures \ref{data} and \ref{smooth} indicate that the shift is more pronounced for  maximums of magnitude of HF-MISO  (in comparison with the maximums of LF-MISO) due to the considerably shorter quantum electron lifetime, $\tau_q^{(3)}$, and, thus, stronger effect of the Dingle factor  in the third subband.  
 
\begin{figure}[t]
\includegraphics[width=1\columnwidth]{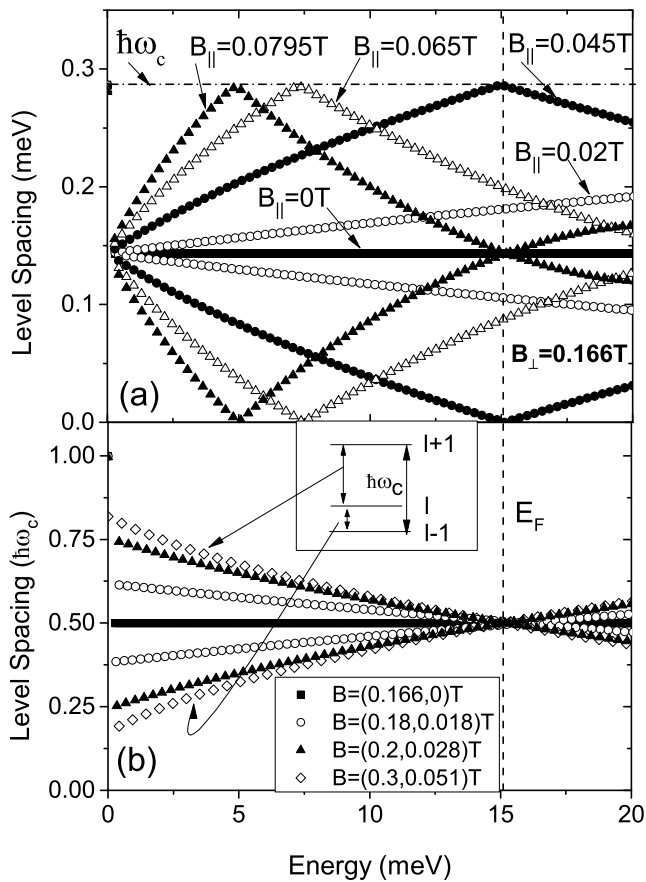}
\caption{(a) Level spacing $\delta E_l=E_{l+1}-E_l$ in the energy spectrum of electrons in fixed $B_\perp$=0.166 T at different in-plane magnetic fields as labeled. At $B_\parallel$=0T the quantum levels are equally spaced with the energy separation $\delta E_l=\hbar \omega_c/2$  producing $k$=3/2 HF-MISO node. At a finite in-plane field the level spacing depends on the energy leading to $k$=2 HF-MISO anti-node at $B_\parallel$=0.045 T and $k$=5/2 HF-node at  $B_\parallel$=0.0795 T;  (b)   Level spacing $\delta E_l=E_{l+1}-E_l$ in the energy spectrum at different  $B_\perp$ and $B_\parallel$ magnetic fields as labeled. These fields correspond to the nodal line between $k$=3/2 and $k$=1/2 HF-nodes shown in Fig.\ref{2D}. The insert explains the meaning of the upper and lower branches of the energy dependence of the level spacing. 
All spectra are obtained at $t_0$=0.215 meV and $d$=36 nm.
}
\label{spectrum}
\end{figure}

Figure \ref{spectrum}(a) presents the difference between energies of $l+1$-th and $l$-th quantum levels of the full electron spectrum obtained in the perpendicular magnetic field $B_\perp$$\approx$0.166 T at  different in-plane magnetic fields as labeled. Each symbol represents a particular level spacing: $\delta E_l=E_{l+1}-E_l$.  At $B_\parallel$=0 and $B_\perp$$\approx$0.166 T the electron spectrum corresponds to the $k$=3/2 HF-MISO node and the LF-MISO minimum. At this node the level spacing $\delta E_l=\hbar \omega_c/2$$\approx$14 meV is the same for all quantum levels except the first two lowest levels, which are separated by $\hbar \omega_c$.  

Due to the complete separation between the lateral and vertical electron motions at $B_\parallel$=0  the level spacing is independent on energy for any $B_\perp$. However in general the level spacing contains two branches corresponding to the nearest upper and lower neighbors of a quantum level. This is shown in the insert to Fig.\ref{spectrum}(b). In the case of a node at $B_\parallel$=0T these two branches coincide everywhere, while for  a node at a finite $B_\parallel$ the two branches intersect in the vicinity of the Fermi energy. The numeric evaluation of the spectrum indicates also that the level spacing does not exceed the cyclotron energy $\hbar \omega_c$ and maintains  the periodicity of the spectrum $\delta E_{l+1}+\delta E_l=\hbar \omega_c$. This is related to the fact that the Hamiltonian $\hat H$ is independent on $k_y$, which preserves the degeneracy of  quantum levels, $g=1/(2\pi l_\perp^2)$,   in in-plane magnetic fields.\cite{hu1992}

Application of an in-plane magnetic field couples the vertical and lateral degrees of freedom. This causes the distribution of level spacing to be energy dependent.  The $B_\parallel$-coupling is due the Lorent's force and, thus, increases with  the electron velocity (energy). At the bottom of a subband the $B_\parallel$-coupling is small and the spectrum is nearly preserved.  At $B_\parallel$=0.02 T the level spacing spreads out almost linearly with the energy.  At a higher in-plane field $B_\parallel$=0.045 T the spread  of the level distribution reaches a maximum $\hbar \omega_c$ in the vicinity of $E_F$. At this condition the two nearest quantum levels coincide with each other. This is the $k$=2 maximum of HF-MISO magnitude and LF-MISO shown in Fig.\ref{2D}.  Further increase of the in-plane field decreases the spread of the level distribution  and in the vicinity of Fermi energy the electron spectrum gradually evolves into a state with nearly uniform level distribution at $B_\parallel$=0.0795 T (intersection of two branches). It corresponds to $k$=5/2 node shown in Fig.\ref{2D}.  At this magnetic field variations of the level spacing is nonlinear with the energy.  

Figure \ref{spectrum}(b) presents  the  level spacing $\delta E_l$ obtained at different perpendicular and  in-plane magnetic fields as labeled.   These fields corresponds to the HF-node $k$=3/2.  The figure shows that an increase of the in-plane magnetic field shifts the $k$=3/2 node to a higher perpendicular magnetic field.  At  small $B_\parallel$ this behavior corresponds to the semi-classical regime and is described below.  

The numerically obtained  evolution of the HF-nodes and LF-maximums in the $B_\perp -B_\parallel$ plane is shown  in Figure \ref{2D}. A good overall agreement between experiment and the theory is found.  A statistical analysis of the experimental and theoretical positions of HF-nodes   indicates the standard deviation below 0.002 T for   the $k$=3/2  HF-MISO node in the range $B_\perp$$ \in$(0.15-0.35) T.  The standard deviation between experiment and  theory in the vicinity of the $k$=1/2 node ($B_\perp$$\in$(0.35-0.5) T ) is found to be significantly  larger (0.02 T). In this region the experimental data deviates  systematically from the theory.  The experimental and theoretical node positions around the $k$=3 region ($B_\perp$$ \in$(0.07-0.2) T ) demonstrate standard deviation below 0.005 T and also deviates systematically from each other  near the apex of the $k$=3 region in the range $B_\perp$$\in$(0.15-0.2) T.   The systematic deviations between the experiment and the theory is discussed below, where we present  different regimes in detail.

\subsection{Semi-classical regime}

\begin{figure}[t]
\includegraphics[width=1\columnwidth]{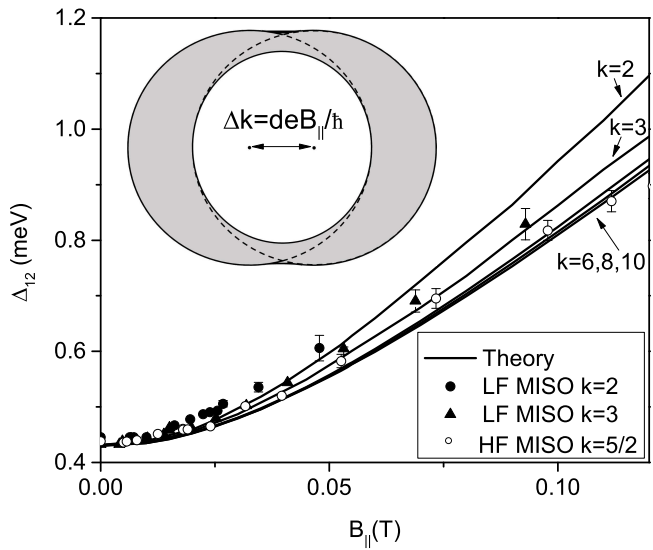}
\caption{ Dependence of the gap $\Delta_{12}$ on in-plane magnetic field extracted from positions of  LF-MISO maximums and a HF-MISO node as labeled. Solid lines represent the gap obtained from the electron spectra evaluated numerically at $t_0$=0.215 meV and $d$=36 nm for different LF-MISO maximums as labeled.   For k=2, 3 and 5/2 standard deviations between experiment and theory are found to be $\delta \Delta_{12}$=0.018, 0.013 and 0.012 meV correspondingly. Sample A. Insert shows semi-classical trajectories in $k$-space at finite in-plane magnetic field $B_\parallel$.}
\label{SC}
\end{figure}
 
The semi-classical regime corresponds to  weak perpendicular magnetic fields at which the Landau-Zener transitions (magnetic breakdown) between different  semi-classical electron trajectories are exponentially weak and  are neglected\cite{hu1992}. Electrons perform semi-classical motion along trajectories corresponding to the symmetric and antisymmetric states. At  $\alpha$=0 ($B_\parallel$=0 T) the semi-classical trajectories are circles with the same origin $\vec k$=0 in the $k$-space. At an energy $E$  the symmetric wave function propagates along the circle with a  radius $k^S>k^{AS}$ and the gap between two subbands $\Delta_{12}$ does not depend on the wave vector $\vec k$.

Application of a parallel field shifts the centers of the two circles by $\delta k=\pm edB_\parallel/2\hbar$ leading to variations of the gap between two subbands with $\vec k$. \cite{bobinger1991,hu1992} The insert to Figure \ref{SC} presents an example of the semi-classical trajectories when a parallel field $B_\parallel$ is applied. The semi-classical trajectory enclosing the gray area corresponds to the symmetric wave function, while the solid line, which is  inside the intersection between two circles, presents the trajectory corresponding to the antisymmetric wave function.  The frequency of quantum oscillations in the reciprocal magnetic field $1/B_\perp$ is proportional to the area enclosed by  semi-classical  trajectory at  an energy $E$.\cite{shoenberg1984,ziman} In the case of HF-MISO the energy $E$ is equal to the energy at the bottom of the third subband: $E=E_3$.  The symmetric state, thus,  has a  frequency $f_{31}$, which is higher than the frequency of quantum oscillations due to the trajectory of the antisymmetric state $f_{32}$. The difference between two frequencies  $f_{12}$  is proportional to the area $A$, shown  in gray  in the insert. In accordance with Eq.(\ref{miso}) at $B_\parallel$=0 the gap $\Delta_{12}=2t_0$ is proportional to  $f_{12}$ and, thus, to the area $A$.  An increase of the in-plane field $B_\parallel$ further shifts  the centers of the two circles increasing the gray area $A$ and, thus, the gap $\Delta_{12}$. 

Figure \ref{SC} demonstrates the increase  of the gap with  in-plane magnetic field. Filled (open) symbols present the gap $\Delta_{12}$ obtained from the relation $\Delta_{12}=k\cdot\hbar\omega_c$, using experimental positions of  LF-MISO maximum (HF-MISO node), where the corresponding index $k$ is an integer (half-integer). Solid lines present the gap obtained from the same relation, using the numerical evaluation of the positions of LF-MISO maximums, which are shown in Fig.\ref{spectrum}(a).  Figure \ref{SC} demonstrates  good agreement between the numerically evaluated gap and experimental data.  The HF-MISO node $k$=5/2 and $k$=3 LF-MISO maximum are clearly seen at high magnetic fields and, thus, the corresponding gaps are  presented in a broader range of  parallel fields in comparison  with the gap obtained from  the  $k$=2 LF-MISO maximum.  The values of experimental and numerical gaps, obtained from the $k$=2 LF-MISO maximum, are found  to be larger than those with higher MISO indexes.   These results are related to  magnetic breakdown of semi-classical trajectories, which is stronger at the $k$=2 LF-maximum.    

Figure \ref{SC} shows also that the curves corresponding to the numerically evaluated gaps collapse at high indexes $k$. This collapse is the signature of the semi-classical regime at which magnetic breakdown is nearly absent and, thus, the obtained gap does not depend on the perpendicular magnetic field.  The strength of  magnetic breakdown is shown in Fig.\ref{2D}.  A comparison between Figures \ref{SC} and \ref{2D} indicates, that for the gap, obtained from the MISO with high indexes $k$,  magnetic breakdown is indeed small at $B_\parallel <$0.1T.  For $k<3$ the probability of  magnetic breakdown increases exceeding 50\%  for $k$=2 at $B_\parallel$=0.1T and $B_\perp$=0.27 T.  

Finally we would like to note that the  dependence of the gap $\Delta_{12}$ on the in-plane  field $B_\parallel$, which is shown in Fig.\ref{SC},  is not the dependence of the difference between the bottoms of the symmetric and anti-symmetric bands. As mentioned above the bottom part of the spectrum is weakly affected by  $B_\parallel$. In contrast to the case of pure perpendicular magnetic field ($B_\parallel$=0T), a finite parallel magnetic field makes the level spacing  energy dependent and the extracted gap represents the relative position of the symmetric and antisymmetric levels in the vicinity of the Fermi energy.

\subsection{High magnetic field regime}

\begin{figure}[t]
\includegraphics[width=1\columnwidth]{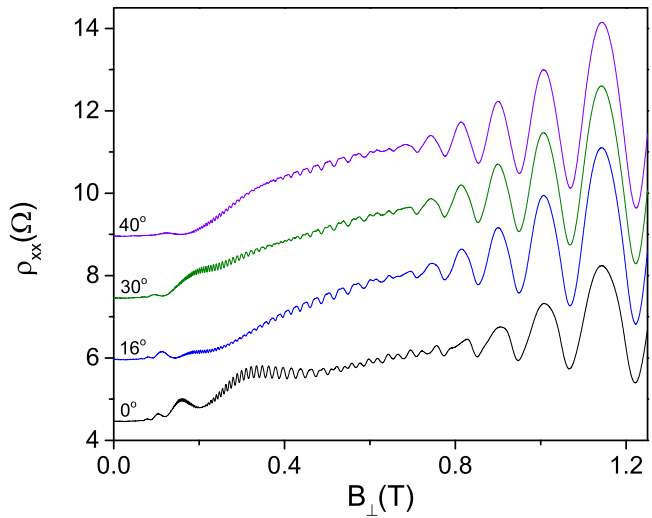}
\caption{Magnetoresitance of GaAs quantum well at different angles between the magnetic field and normal to the sample. Three upper curves are shifted for clarity.  T=4.2K. Sample B.}
\label{HFa}
\end{figure}

In a strong magnetic field $B_\perp$ the cyclotron energy exceeds the  gap: $\hbar \omega_c \gg \Delta_{12}$.  In this range of magnetic fields Shubnikov de Haas (SdH) oscillations are well developed. Figure \ref{HFa} presents  the magnetoresistance taken at different angles $\alpha$ between the direction of the applied magnetic field and the normal to the 2D sample.   At temperature $T$=4.2K  SdH oscillations appear in  $B_\perp$ exceeding 0.5T. At a smaller field these oscillations are significantly damped and only MISO are observable at $T$=4.2 K.\cite{dietrich2015} 

\begin{figure}[t]
\includegraphics[width=1\columnwidth]{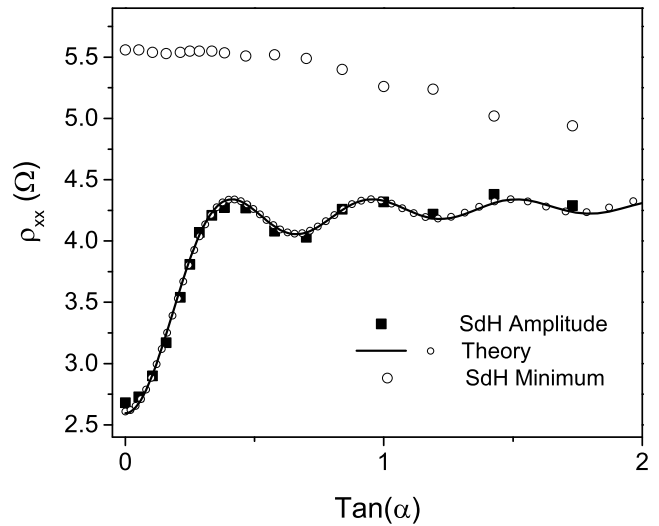}
\caption{Dependence of the swing of SdH oscillations at $B_\perp$=1.14 T on $tan(\alpha)$. Filled squares present experimental data obtained from the magnetoresitance curves shown in Fig.\ref{HFa} with an accuracy approximated by the size of the symbols. Solid line (small open circles) is a theoretical dependence obtained from numerical (analytical) evaluation of the electron spectrum at fixed  $B_\perp$=1.14 T and different $B_\parallel$ corresponding to different angles $\alpha$ using $A_{SdH}$=2.17 $\pm$0.01 Ohm and $\gamma$=0.479 $\pm$0.01 meV as fitting parameters and $t_0$=0.215 meV,  $d$=36 nm and $E_F$=15.1meV. Big open circles present the angle dependence of the sample resistance in the SdH minimum at $B_\perp$=1.07 T. T=4.2K. Sample B. }
\label{HFb}
\end{figure}

The amplitude of  SdH oscillations increases considerably  with  angle. Figure  \ref{HFb} demonstrates the angular dependence of a swing (doubled amplitude) of SdH oscillations taken at $B_\perp$=1.14T.   The swing of SdH oscillations is measured between upper and low branches of the envelope of SdH oscillations. The upper (low) branches of the envelope are obtained using a cubic spline between maximums (minimums) of SdH oscillations.\cite{dietrich2015} The swing of  oscillations increases monotonically from 2.65 $\Omega$ at $tan(\alpha)$=0 to about 4.35 $\Omega$ at $tan(\alpha)$$\approx$0.4. Then the oscillation swing demonstrates  small periodic variations with $tan(\alpha)$.  

To evaluate the SdH amplitude we used the following expression. The SdH amplitude depends on the level spacing $\delta E_l$  in the vicinity of the Fermi energy since all quantum states below $E_F$ are completely occupied. This fact allows for a modification of  the actual distribution of the occupied levels inside  subbands to simplify the mathematical description of  SdH oscillations.  Below we use  a level distribution with equal spacing, $\hbar \omega_c$, inside each subband. The two periodic sets of levels  are shifted with respect to each other by the value corresponding to the actual spacing $\delta E_l$ between the nearest quantum levels in the vicinity of the Fermi energy.  The spectrum modification doesn't change the number of occupied states preserving the total electron density.  The modified spectrum is similar to the spectrum  at $B_\parallel$=0 T with $\Delta_{12}=\delta E_l$ and yields SdH oscillations approximated  by  a  cosine function:
\begin{align}
\Delta\rho_{SdH}^{(i)}=A_{SdH}^{(i)} \cos\left(\frac{2\pi(E_F-E_*^{(i)})}{\hbar\omega_c}\right) 
\label{sdh}
\end{align} 
The SdH amplitude, $A_{SdH}^{(i)}$,  includes Dingle factor $d_i=exp(-\pi/\omega_c \tau_q^{(i)})$ and a temperature damping factor $A_T=x/sinh(x)$, where $x=2\pi^2kT/\hbar \omega_c$.\cite{shoenberg1984}   In contrast to  HF-MISO the phase of the cosine contains the Fermi energy instead of the energy of the bottom of the third subband $E_3$ (see Eq.(\ref{miso})). The energy $E_*^{(i)}$ corresponds to the bottom of the modified spectrum of the $i$th subband.  
Due to the nearly equal quantum scattering times in the symmetric and antisymmetric subbands both SdH oscillations  have the same amplitude $A_{SdH}$.  The sum of the two oscillations $\Delta \rho_{SdH}=\Delta \rho_{SdH}^{(1)}+\Delta \rho_{SdH}^{(2)}$ can be  presented as a product of two cosines:
\begin{align}
\Delta \rho_{SdH}&=2A_{SdH} \cos\left(\frac{\pi E_+}{\hbar\omega_c}\right)\cos\left(\frac{\pi E_-}{\hbar\omega_c}\right)    \nonumber\\ 
&\approx 2A_{SdH}\cdot \cos\left(\frac{\pi E_+} {\hbar\omega_c}\right)\left(1- \frac{1}{2} \left(\frac{E_-}{\gamma}\right)^2\right),
\label{beating}
\end{align}
where $E_+=2E_F-E_*^{(1)}-E_*^{(2)}$ is the sum and $E_-=E_*^{(2)}-E_*^{(1)}$ is the difference between the energy terms in Eq.(\ref{sdh}).   The energy $E_+$ describes the high frequency content of the SdH oscillations, which is intact since both the total electron density and the Landau levels degeneracy  $g=1/(2\pi l_\perp^2)$ are preserved in the modified spectrum.  We note also that the difference between the terms  equals  the actual level spacing near $E_F$:  $E_-=\delta E_l$.  Thus Eq.(\ref{beating}) provides a description of SdH oscillations corresponding to the actual spectrum $E_l$. Since in  high magnetic fields the cyclotron energy  is  considerably higher than  the level spacing $\delta E_l$ the low frequency cosine, modulating  the SdH amplitude, is approximated  by a Tailor series.   At $B_\perp$=1.14 T the factor $\gamma=\hbar \omega_c/\pi$$\approx$0.63 meV is larger than $\Delta_{12}$=0.43 meV at $B_\parallel$=0T.

The approximation of SdH oscillations by a single cosine is valid when the swing of SdH oscillations is small in comparison with the Drude resistance at $B_\perp$=0 T. In the studied case the oscillation swing is comparable with the Drude resistance and,  thus,  higher harmonics  of SdH oscillations  should be accounted for. In the case of a small level spacing between  subbands : $E_- \ll \hbar \omega_c$,  variations of the amplitude of the higher harmonics  with the angle $\alpha$ are expected to be also proportional to $E_-^2$ similar to the variations of the fundamental harmonic in Eq.(\ref{beating}). Taking this into account  we  compare the theory and experiment  using  Eq.(\ref{beating}) with $A_{SdH}$ and $\gamma$ as  fitting parameters.  

Shown in Fig.\ref{HFb} the solid line presents the angular dependence of the swing of SdH oscillations yielded by Eq.(\ref{beating}). The energy $E_-$ is extracted from the electron spectra evaluated numerically at fixed $B_\perp$=1.14T and different  $B_\parallel=B_\perp tan(\alpha)$. For each combination of $B_\perp$ and $B_\parallel$ the energy spectrum $E_l$ is computed with the same model parameters $t_0$=0.215 meV and $d$=36 nm used in previous spectrum computations shown Fig.\ref{2D}-\ref{SC}. The standard deviation between the experimental data and the numerical evaluation of the SdH amplitude is found within 0.05$\Omega$ indicating a good agreement between the experiment and the proposed model.

Shown in Fig.\ref{HFb} the small open circles present a theoretical dependence obtained from the analytical expression for the level spacing in high magnetic fields: $\delta E_N=2t_0exp(-\theta) L_N(2\theta^2)$, where $\theta=B_\parallel d/(2B_\perp l_\perp)$ .\cite{hu1992} At high $N$ the Laguerre function, $L_N(x)$, is approximated by a Bessel function, $J_0(x)$, yielding the level spacing $E_-=\delta E_l$$\approx$$2t_0J_0[k_Fdtan(\alpha)]$, where $k_F=(2mE_F)^{(1/2)}$ is wave number at the Fermi energy. The analytical evaluation of the swing of SdH oscillations demonstrates better agreement with the numerical data yielding the standard deviation within 0.01$\Omega$. The results indicate that the inaccuracy of the numerical computations of the electron spectrum is likely not the main source of the deviations between the experiment and theory.  

Fig.\ref{HFb}  demonstrates oscillations and  the complete reduction of the tunneling magnitude in the maximums of the oscillations. At an angle $\alpha_n$ corresponding to  $n$-th maximum, the beating pattern between two SdH oscillations is absent since the beating period ($\sim1/E_-)$ is infinite at the angle $\alpha_n$. The absence of the beating pattern at  "magic angles" as well as the beating of SdH oscillations  is observed   in strongly anisotropic layered organic materials.\cite{amro1,amro2} These resistance oscillations   with the angle $\alpha$ in high magnetic fields  have been seen recently in double quantum wells in the   Quantum Hall effect regime.\cite{gusev2007,gusev2008}. 

The evolution of the level spacing with the angle $\alpha$  can be understood using an intuitively appealing picture of the phenomenon.\cite{moses1999,yakov2006} In the bilayer geometry the tunneling between  layers $a$ and $b$ can be described by the Hamiltonian 
\begin{align}
H_t=t_0 \int \phi_a^*({\bf r}) \phi_b({\bf r})exp(ieA_z({\bf r})d/\hbar)d^2r+H.c.,
\label{tun1}
\end{align}
where vector potential $A_z=B_\parallel x$ corresponds to the in-plane magnetic field directed along $y$-axes: $\vec B=(0,-B_\parallel,0)$. In the presence of $B_\perp$, an electron with Fermi energy  propagates along the cyclotron orbit with radius $r_c$.  The gauge phase in Eq.(\ref{tun1}) oscillates  along the electron trajectory leading to a modification of the tunneling. The effective tunneling amplitude $t$ is obtained by the phase averaging\cite{moses1999}:
 \begin{align}
t=t_0\langle exp(ieB_\parallel x(t)d/\hbar)\rangle_t=t_0J_0(k_Fd  tan(\alpha )).
\label{tun1}
\end{align}      
The brackets represent a time average over the period of the cyclotron motion and $x(t)=r_ccos(\omega_c t)$ is the $x$-coordinate of the electron.  The obtained expression coincides with the one used for fitting experiment data in Fig.\ref{HFb}. 

In Fig.\ref{HFb} large open circles represent the dependence of the resistance, $R_{min}$, in the SdH minimum at $B_\perp$=1.07 T  on the  in-plane magnetic field. The notable feature of the observed behavior is the  stability of the resistance value in a broad range of  $\alpha$ despite  the significant variations of the SdH amplitude in the same angular range.  The stability is found for all other SdH minimums shown in Fig.\ref{HFa}. The resistance $R_{min}$ starts to decrease with the angle after the level spacing $E_-$ reaches the first maximum at a finite $B_\parallel$.  In contrast to the SdH amplitude the $R_{min}$  depends on the behavior of the non-oscillating background which was beyond the scope of this paper.

Finally we would like to mention that  in the studied samples SdH oscillations are comparable with MISO near  $B_\perp$$\approx$0.5T in the vicinity of the HF-MISO node $k$=1/2 shown in Fig.\ref{2D}.\cite{dietrich2015} The presence of SdH oscillations may affect the position of this node since  the phase of  SdH oscillations is shifted by $\pi$ with respect to the phase of MISO \cite{leadley1992,raikh1994,sander1998,dietrich2015}. An  analysis of  the beating of quantum oscillations indicates that the nodes of both SdH oscillations and HF-MISO occur at the same magnetic field. In accordance with numerical computation the node $k$=1/2 occurs at $B_\perp$=0.5T.  If at $B_\perp$=0.5T the SdH amplitude is larger than the amplitude of HF-MISO,  then the  SdH oscillations  dominate at $B_\perp>$0.5 T since the oscillations grow faster than MISO due to the  additional temperature factor $A_T(B_\perp)$ (see Eq.(\ref{sdh})).  At $B_\perp<$0.5 T the SdH oscillations are comparable with HF-MISO and  the destructive interference  between two oscillations reduces the overall oscillation amplitude. It makes the actual node  to be broader and shifted toward  smaller magnetic fields. This is  indeed seen in Fig.\ref{2D} near the $k$=1/2 node.  We suggest that the systematic deviation between experiment and the theory observed at the $k$=1/2 node is the result of  destructive interference between SdH oscillations and MISO.
   
\subsection{Magnetic breakdown regime}
As mentioned above an application of parallel field, $B_\parallel$,  shifts the centers of  cyclotron orbits in two layers  by $\delta k=\pm edB_\parallel/2\hbar$ leading to variations of the gap between the  two subbands with $\vec k$. \cite{bobinger1991,hu1992} The smallest  gap  occurs  in a small  region with  lateral size  $\Delta k_0(B_\parallel)$ near the intersections between  two circles shown in the insert to Fig.\ref{SC}.   At small magnetic fields electrons circulate along  the semi-classical trajectories $\vec k_S(t)$ and $\vec k_{AS}(t)$,  corresponding to the symmetric and antisymmetric states.  The probability of  magnetic breakdown  between these  trajectories  depends strongly on the time $\Delta t$ during  which  electrons pass the region  with  the smallest gap: $\Delta t \sim \hbar \Delta k_0/eV_FB_\perp$.   At small magnetic fields the time $\Delta t \gg \hbar/t_0 $ is long enough to establish the gap between subbands and  magnetic breakdown  is exponentially suppressed.\cite{cohen1961,blount1962,slutskin1968}.  An increase  of both $B_\perp$ and $B_\parallel$ increases the probability of magnetic breakdown. In a WKB approximation an expression has been obtained for the breakdown probability $P_{MB}$\cite{hu1992}:
\begin{align}
P_{MB}=exp(-\omega^*_c/\omega_c),
\label{break1}
\end{align}      
where
\begin{align}
\omega^*_c=\frac{\pi t_0^2}{E_F(Q/k_F)[1-(Q/2k_F)^2]^{1/2}}.
\label{break2}
\end{align} 
Here $Q=deB_\parallel / \hbar$ is the relative displacement of the two Fermi circles due to $B_\parallel$.

The 50\% probability of the magnetic breakdown  at different $B_\perp$ and $B_\parallel$ is plotted in Fig.\ref{2D} for sample A. Fig.\ref{2D} demonstrates a  correlation  of  magnetic breakdown with the behavior of the nodal lines. In particular  the  collapse of 5/2 and 7/2 nodes occurs at a higher $B_\parallel$ than the one of  1/2 and 3/2 nodes  that  is  in  qualitative agreement with the behavior of the line describing  magnetic breakdown.  

The notable feature of magnetic breakdown is the growth of quantum oscillations  with frequency equal to  half sum of the frequencies corresponding to symmetric and antisymmetric semi-classical trajectories: $f_+=(f_{31}+f_{32})/2$.\cite{hu1992,harff1997,shabani2008} The frequency,   $f_+$, is due to the circular orbital motion of an electron completely located in one of the  layers.  The following consideration helps to understand the origin of the frequency $f_+$.  In accordance with Eq.(\ref{break1}) at  small $B_\perp$: $\omega_c \ll \omega_c^*$ the probability of  magnetic breakdown is exponentially small and, thus, can be neglected.  In the absence of  magnetic breakdown  electrons follows the semi-classical trajectories and the spectrum of the quantum oscillations contains frequencies $f_S=f_{31}$ and $f_{AS}=f_{32}$ corresponding to the symmetric and antisymmetric subbands. An example of the semi-classical  trajectories corresponding to the two subbands  in a finite $B_\parallel$ is shown in the insert to Fig.\ref{SC}.  Following  the semi-classical trajectory  an electron moves periodically between the top and bottom layers.  An increase of the perpendicular magnetic field  enhances   the probability of   magnetic breakdown.  At $\omega_c > \omega_c^*$ the electron has a considerable probability to cross the tunneling gap and to follow a trajectory, which is not perturbed by the tunneling. This trajectory is a circular orbit located completely in a single layer. These orbits are presented by dashed lines in the insert to Fig.\ref{SC}. The insert indicates that the total area of the two circles equals  the sum of the area inside the perimeter of the shifted circles (symmetric subband) and the area of the overlap of the two circles (antisymmetric subband). Since the frequencies of the quantum oscillations are proportional to the corresponding areas\cite{shoenberg1984} the relation between different areas  yields:   $2f_+=(f_{31}+f_{32})$. 

\begin{figure}[t]
\includegraphics[width=0.9\columnwidth]{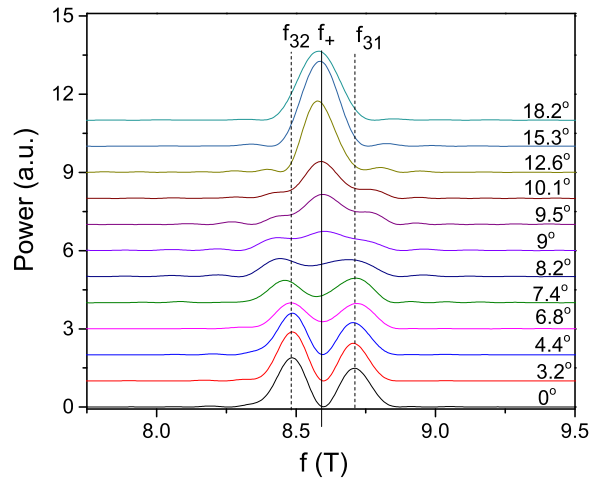}
\caption{Fourier power spectra of  MISO at different angles as labeled. The spectra are obtained in the interval of reciprocal magnetic fields between 2 and 10 1/T shown in Fig.\ref{smooth}. The spectra are vertically shifted for clarity. Sample A. }
\label{MB}
\end{figure}

Figure \ref{MB} shows the increase of the amplitude of  quantum oscillations with  frequency $f_+$ as  magnetic breakdown increases.  At zero angle $\alpha$  magnetic breakdown is absent and the spectrum of  MISO contains only two frequencies $f_{31}$ and $f_{32}$ corresponding to symmetric and antisymmetric subbands. With an increase of  $\alpha$ the magnitude of  parallel magnetic field and, thus, the probability of  magnetic breakdown  increase. The enhanced magnetic breakdown decreases  the magnitude of  MISO  and increases the magnitude of the oscillations corresponding to  the  isolated 2D layers, which appear at  frequency  $f_+ \approx (f_{31}+f_{32})/2$. At   $\alpha>$13$^o$ the oscillations at frequency $f_+$ are predominant.  

Figure \ref{MB} demonstrates also an increase of the difference between frequencies $f_{31}$ and $f_{32}$ corresponding to symmetric and antisymmetric subbands with the angle $\alpha$.  The increase of $\Delta f =f_{31} - f_{32}$ is related to the increase (decrease) in size of the symmetric (antisymmetric) orbits with  the increase of  in-plane magnetic field\cite{bobinger1991}.  

Magnetic breakdown is  the origin of the  collapse of  HF-MISO nodes and the nodal confinement  of  LF-MISO with odd indexes $k$ as shown in Fig.\ref{2D}.     To understand this relation we note that the phase of the oscillations with frequency $f_+$ is the same as the phase of HF-MISO for even $k$ and is shifted by $\pi$ for odd $k$. Fig.\ref{smooth} shows this correspondence:  in the $k$=1 region  maximums of HF-MISO   at $\alpha$=0$^0$ (no magnetic breakdown) correspond to minimums of HF-MISO at $\alpha$=15.3$^o$ (strong magnetic breakdown), while in the $k$=2 region these two HF-MISOs are in-phase.   Magnetic breakdown  admixes  oscillations similar to one at $\alpha$=15.3$^o$ to the oscillations at $\alpha$=0$^o$, and, therefore,  decreases the magnitude of the oscillations corresponding to the odd $k$.  Thus, the magnitudes of odd $k$ HF-MISO  and corresponding LF-MISO maximum decrease.  

\begin{figure}[t]
\includegraphics[width=1\columnwidth]{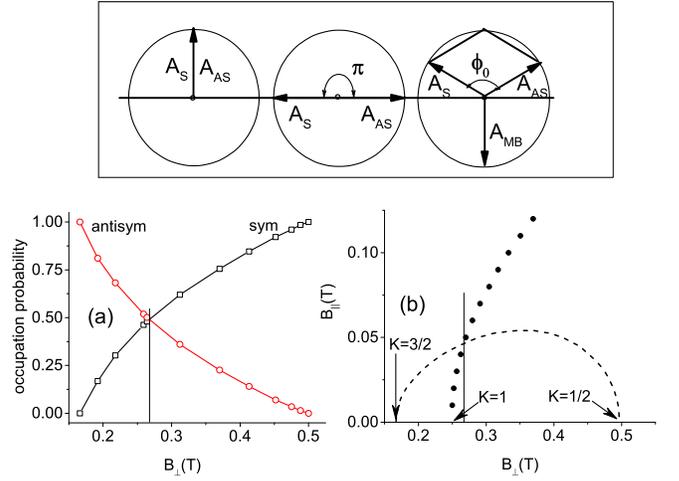}
\caption{(a) Open symbols present  contributions of symmetric and antisymmetric states to population of  quantum level $\vert l\rangle =c_S\vert S \rangle+c_{AS}\vert AS\rangle$ in the vicinity of Fermi energy at different  $B_\perp$ and $B_\parallel$ corresponding  to the nodal line shown in (b). The eigenstate $\vert l\rangle$ of the Hamiltonian $\hat H$ is computed numerically; (b) Dashed line presents nodal line enclosing the $k$=1 region. Solid symbols show positions of  level   $\vert l\rangle =(\vert S \rangle+\vert AS\rangle)/\sqrt2$ with the  equal population of the symmetric and antisymmetric states in $B_\perp-B_\parallel$ plane.  The dependence intersects the nodal line at  the same  magnetic field $B_\perp$ at which two lines shown in (a) intersect. Sample A. Insert shows phasor diagram describing interference of MISOs presented by $\vec A_S$ and $\vec A_{AS}$ with quantum oscillations,  $\vec A_{MB}$, induced  by magnetic breakdown.}
\label{nodal}
\end{figure}

The mixing moves the nodes, confining an odd $k$ region,  toward each other. The insert to Fig.\ref{nodal} presents a phasor diagram illustrating this property.  Eq.(\ref{miso}) describes MISOs corresponding to symmetric and antisymmetric subbands by  cosine functions with frequencies $f_{3i} \sim \Delta_{3i}$. Shown in the insert two vectors $\vec A_S$ and $\vec A_{AS}$ represent the amplitude and phase of the two cosine functions  corresponding to symmetric  and antisymmetric  subbands.   Without  magnetic breakdown the two oscillations are  in phase and, thus, the   two vectors  are in the same direction at HF-MISO antinodes.  Below we  consider  the $k$=1 region. As shown in Fig.\ref{smooth} the $k$=1 antinode occurs at   $1/B_\perp^{an}$$\approx$4 1/T.   A right shift of the $1/B_\perp$ to the nearest node $k$=3/2, located at $1/B_\perp^{n}$$\approx$6 1/T,  destroys the parallel alignment  between the two vectors. In  a reference frame rotating with frequency $f_+$ the right shift rotates the vector $\vec A_S$ ($\vec A_{AS}$) counter clockwise (clockwise) yielding a phase angle $\pi$ between two vectors, that corresponds to an orientation of two vectors in opposite directions. At the node the sum of the vectors is zero that  corresponds to  the completely destructive interference between the two oscillations.  

Magnetic breakdown adds an additional vector, $\vec A_{MB}$, to  the phasor diagram. The amplitude of $\vec A_{MB}$ corresponds to the amplitude of the quantum oscillations at frequency $f_+$. To simplify the presentation we use the magnitude of the vector $\vec A_{MB}$ to be the same as the other magnitudes.  In the rotating frame the vector $\vec A_{MB}$ is oriented  down since in   odd $k$ regions the phase of oscillations, induced by magnetic breakdown, is shifted by $\pi$ with respect to the MISO phase at the antinode.  In the magnetic breakdown regime the node  occurs at a phase difference $\phi_0$ between $\vec A_S$ and $\vec A_{AS}$, at which the sum of three vectors $\vec A_S+\vec A_{AC}+\vec A_{MB}$ is zero. The angle $\phi_0$ is smaller than $\pi$ and, thus, corresponds to a node located  at $1/B_\perp^{MB}<1/B^{n}$ closer to the antinode position at $1/B_\perp^{an}$.  At a larger magnitude $A_{MB}$ the angle $\phi_0$ is smaller  indicating  further displacement of the node position toward the antinode.   A similar consideration of  the $k$=1/2 node shows the node displacement in the opposite direction i.e. again toward the antinode at $1/B_\perp^{an}$$\approx$4 1/T. Finally at $\vert \vec A_{MB}\vert=\vert \vec A_ S\vert+\vert \vec A_{AS}\vert$ the phase difference $\phi_0$=0 and the  two nodes collapse   bounding completely the  $k$=1 region in the $B_\perp-B_\parallel$ plane. 

Below we consider  additional  properties  of the $k$=1 region and the nodal line between $k$=1/2 and $k$=3/2 nodes. Figure \ref{nodal} demonstrates the probabilities of the   population of the symmetric and antisymmetric states along the nodal  line for a quantum  state $\vert l\rangle$ in the vicinity of the Fermi energy. The probabilities are obtained from an analysis of the eigenvectors of  the Hamiltonian $\hat H$ (see Eq.(\ref{diag})  and Eq.(\ref{offdiag})). The numerical computations indicate that at the nodal line  the  eigenvector $\vert l\rangle$ contains primarily the contributions from one symmetric and one antisymmetric state: $\vert l\rangle \approx c_S\vert S \rangle+c_{AS}\vert AS\rangle$, where $c_A$ and $c_{AS}$ are the amplitude of the  states.  All other contributions to the level population are within a few percents and are neglected.  

Fig.\ref{nodal}(a) presents the probability $P_S=c_S^2$ and $P_{AS}=c_{AS}^2$ at different $B_\perp$ and $B_\parallel$  corresponding to the nodal line around the $k$=1 region. The figure demonstrates that  at the node $k$=3/2 located at  $B_\perp$=0.166 T and  $B_\parallel$=0 T the quantum state  $\vert l\rangle$ is completely antisymmetric: $c_S$=0 and $c_{AS}$=1. This node is due to the interlayer tunneling only.  A  shift along the nodal line increases both  $B_\perp$ and  $B_\parallel$ enhancing the magnetic breakdown, which in  turn  increases  (decreases) the population of the symmetric (antisymmetric) states.  At $B_\perp$=0.268 T the two states  are equally populated and  $\vert l\rangle =(\vert S \rangle \pm \vert AS\rangle)/\sqrt 2$.  Fig.\ref{nodal}(b) indicates that an approach of  $B_\perp$ to the $k$=1/2 node decreases  $B_\parallel$ and  the magnetic breakdown. Finally at the node $k$=1/2 located at $B_\perp$=0.5 T and  $B_\parallel$=0 T the probability of  magnetic breakdown is zero and the state $\vert l\rangle$ is formed again by the interlayer tunneling only. However in contrast to the state $\vert l\rangle$ at $k$=3/2  the state $\vert l\rangle$  is now completely symmetric.  The transformation of the state symmetry occurs while the state $\vert l\rangle$ was always gapped since at a nodal line the energy levels are evenly spaced by $\hbar \omega_c/2$.  

The observed smooth transformation of the level symmetry   is due to a  repulsion of the quantum levels induced by  magnetic breakdown.   Without the magnetic breakdown at $B_\parallel$=0T the symmetry of the state $\vert l\rangle$  changes abruptly with the perpendicular magnetic field at $B_\perp$=0.25 T corresponding to the $k$=1 LF-MISO maximum. At this magnetic field the energies of $\vert N+1,S\rangle$ and $\vert N,AS\rangle$ states of the symmetric and antisymmetric bands  coincide and, thus, the gap between these levels is zero.  At $B_\perp$=0.25 T and $B_\parallel$=0T these two levels cross each other.  Magnetic breakdown opens up a gap between the levels  leading to a  smooth transformation of the symmetry of the eigenvector $\vert l\rangle$.  The solid symbols show  locations of the quantum level with  equal symmetric and antisymmetric population   $\vert l_{eq}\rangle =(\vert S \rangle+\vert AS\rangle)/\sqrt2$. The line divides the area under the nodal line on the symmetric and antisymmetric parts. At high $B_\parallel$ the location of the level $\vert l_{eq}\rangle$ approaches the location of the $k$=2 LF-MISO maximum (not shown).
 
We note that near $B_\perp$=0.25 T and $B_\parallel$=0T the numerical simulations show a substantial  increase of the level splitting for small magnetic breakdown between the two states indicating a strong sensitivity of the electron spectrum to the parallel magnetic field  at $k$=1. Such strong sensitivity of the spectrum to the $B_\parallel$ is also seen in the perturbation expansion of the spectrum vs $B_\parallel d/2B_\perp l_\perp$.\cite{hu1992} Eq.(4.7) of the paper\cite{hu1992}  indicates a divergence of the second order correction to the level spacing $\delta E_l$ at $\hbar \omega_c=2t_0$ corresponding to the $k$=1 MISO maximum. These results agree with the presented experiments demonstrating  significant sensitivity of the MISO maximum at $k$=1 to  in-plane magnetic field.  

Finally we would like to discuss the  discrepancy between experimental and theoretical positions of the HF-MISO node, which is observed near the apexes of  the odd regions $k$=3 and $k$=5, where nodal lines $(k\pm 1/2)$ meet each other in Fig.\ref{2D}. This discrepancy is not related to SdH oscillations since the SdH amplitude is negligibly small at these magnetic fields\cite{dietrich2015}. Figure \ref{2D} demonstrates that the probability of the magnetic breakdown near  the apexes is  50\%.  Numerical computations reveal that, near these apexes  the eigenstate of the studied Hamiltonian $\hat H$ contains comparable contributions from symmetric  and antisymmetric quantum states of many Landau levels. Thus the quantum state possesses  a complex set of semiclassical trajectories.       In general different trajectories provide different contributions to the transport\cite{pippard1962,pippard1964,slutskin1968,hu1992}. This property of the quantum states has not been  taken into account in the presented model. We suggest that the observed deviations between experiment and theory are related to the complex structure of  quantum  levels near the apexes of odd $k$ regions. The complex structure  is induced by  magnetic breakdown. 

\section{Conclusion}
Magneto-inter-subband oscillations of the resistance of two dimensional electrons are investigated  in wide GaAs quantum well with three populated subbands placed in tilted magnetic fields. At zero in-plane magnetic field the oscillations demonstrate three distinct frequencies $f_{ij} \sim \Delta_{ij}$ in reciprocal perpendicular magnetic field $1/B_\perp$.  The low frequency oscillations, LF-MISO,  is due to  enhancement of the electron scattering when Landau levels of  two lowest, symmetric and antisymmetric subbands, are aligned with each other. These oscillations obey the relation: $\Delta_{21}=k\cdot \hbar \omega_c$.  Related to the third subband two HF-MISOs  have much higher frequencies: $f_{31}$ and $f_{32}$ due to the higher energy difference between bottoms of the third and lowest subbands: $\Delta_{3i} \gg \Delta_{21}$. HF-MISOs demonstrate a distinct beating pattern with a beat frequency $f_{beat}=(f_{31}-f_{32})/2$.    A rotation of the direction of the magnetic  field by an angle $\alpha$ from the normal to the samples produces dramatic  changes of MISO. At  small $\alpha$ the LF-MISO maximum and the corresponding antinode of HF-MISO at $k$=1 disappear. In the $B_\perp-B_\parallel$ plane the $k$=1 region  is found to be bounded by a continuous nodal line connecting the $k$=3/2 and $k$=1/2 nodes of HF-MISO. Similar nodal bounding is found for other odd $k$ regions. This  bounding  correlates with the probability of  magnetic breakdown, $P$, between semi-classical trajectories corresponding to symmetric and antisymmetric subbands. The nodal bounding  is mostly completed at $P<$1/2 for $k$=1 and $k$=3 regions. The Fourier analysis of the oscillations  beyond the bounded regions  shows the dominant contribution of the oscillations to be of  period  $f_+=(f_{31}+f_{32})/2$ corresponding to the electron orbits located at either side of the quantum well and populated by  magnetic breakdown.  The location of the HF-MISO nodes as well as the evolution of the LF-MISO maximum on $B_\perp-B_\parallel$ plane are found to be in an excellent agreement with numerical evaluations of the electron spectra. 

Authors thank Scott Dietrich for help with experiments.  This work was supported by the National Science Foundation (Division of Material Research - 1104503),  the Russian Foundation for Basic Research (project no.16-02-00592) and  the Ministry of Education and Science of the Russian Federation.


\begin{thebibliography} {}
\bibitem{deviation} The standard deviation $\delta y$ between two correlated data sets, $y_i$ and $y_i^0$   is computed using the following formula: $\delta y=(\sum_{i=1}^N (y_i-y_i^0)^2/N)^{1/2}$, where  $N$ is number of data points in each set and integer $i$ correlates the data.
\bibitem{shabani2008} A similar frequency is observed in paper \cite{shabani2}.  The frequency is related to a magnetic breakdown between heavy and light heavy holes, which is induced by a mechanical strain applied to GaAs samples.
\end{thebibliography}
\end{document}